\documentclass[final,5p,times,twocolumn,authoryear]{elsarticle}

\usepackage{amssymb}
\usepackage{amsmath}
\usepackage{lipsum}
\usepackage{mathrsfs}

\journal{Mod. Phys. Lett. A}

\begin{document}

\begin{frontmatter}




\title{Non-degenerate and degenerate wormholes:\\a unified approach}
\author{Juri Dimaschko}
\address{Technische Hochschule Lübeck, Mönkhofer Weg 239, 23562 Lübeck, Germany}

\begin{abstract}
A generalized notion of degenerate wormholes is introduced,  defined by the vanishing of the metric determinant \(g\) at the throat. It is described by the polynomial, \(g^{2}\)-modified Einstein field equations. Building on this framework, we show that both the Einstein–Rosen bridge and the Klinkhamer defect wormhole are exact vacuum solutions of the \(g^{2}\)-modified equations, valid globally including at the degenerate throat, while the Klinkhamer configuration additionally admits traversable geometries with \(b>2M\), where \(b\) sets the length scale of the wormhole throat and  \(M\) is a mass parameter. In contrast, standard Morris–Thorne and thin-shell wormholes, governed by the conventional (non‑regularized) Einstein equations, are intrinsically non‑degenerate and necessarily supported by exotic stress–energy. Within a unified regularized system with matter, both thin‑shell and Klinkhamer wormholes appear as two qualitatively distinct classes of states—non‑degenerate with exotic matter versus degenerate with vacuum—sharing the Einstein–Rosen bridge as a common limiting configuration. This unified viewpoint clarifies why classical null‑energy‑condition no‑go theorems apply only to the non‑degenerate sector and suggests the possibility of stationary degenerate traversable wormholes that do not require NEC violation.
\end{abstract}




\end{frontmatter}

\section{Introduction}

Over the past 35 years, a broadly accepted viewpoint has emerged in the study of wormholes: within classical general relativity, traversable wormhole geometries are topologically admissible but require exotic matter that violates the null energy condition (NEC) in order to exist or remain stable \citep{MT,visser,hochberg}. This conclusion has been established for a wide variety of spherically symmetric and more general wormhole solutions constructed as non‑degenerate metrics solving the standard Einstein field equations with matter.

A less explored question is whether this restriction is fundamentally geometric, or instead reflects the implicit assumption that the spacetime metric is everywhere non‑degenerate. The usual formulation of Einstein’s equations presupposes a pseudo‑Riemannian metric \(g_{\mu\nu}(x)\) with nonvanishing determinant \(g(x)=\det\|g_{\mu\nu}\|\neq 0\); metrics with \(g(x)=0\) lie outside its domain of applicability. As a result, essentially all classical results on wormhole existence and NEC violation pertain only to this non‑degenerate sector of configuration space.

By contrast, degenerate wormhole geometries—where \(g(x)\) vanishes on the throat—are naturally described within the Einstein–Rosen “polynomial” formulation, in which the field equations are multiplied by \(g^{2}\) and become regular for both non‑degenerate and degenerate metrics \citep{einstein1,peres,katanaev}. In particular, the Einstein–Rosen bridge and the more recent Klinkhamer defect wormhole provide explicit examples of such degenerate configurations that solve the \(g^{2}\)-modified vacuum equations and do not require explicit matter sources \citep{klinkhamer1,klinkhamer2,klinkhamer3,wang}.

These defect wormholes have attracted considerable attention because, from the standpoint of the standard Einstein equations, their degenerate throat obstructs the computation of the curvature tensor and appears to evade the usual NEC‑violation arguments. Critics have therefore argued that Klinkhamer’s construction is at best a formal vacuum solution valid only outside the throat and that, once the throat is treated properly, one simply recovers the familiar thin‑shell wormhole supported by exotic matter \citep{feng,baines}. This raises a basic conceptual issue: do non‑degenerate thin‑shell wormholes and degenerate defect wormholes actually represent the same physical state expressed in different coordinates, or do they belong to distinct classes of states governed by different forms of the Einstein field equations? 

The purpose of the present work is to place non‑degenerate and degenerate wormholes on the same footing by introducing a relativistically invariant classification based on the degeneracy of the metric determinant \(g\), and by formulating a unified \(g^{2}\)-modified Einstein–matter system that encompasses both the usual Morris–Thorne type wormholes and defect wormholes of Einstein–Rosen/Klinkhamer type. 

The paper is structured as follows.

In Sec. 2 we contrast two complementary constructions—the Thorne approach, which starts from a two‑sheeted Morris–Thorne ansatz and standard Einstein equations with matter, and the Einstein–Rosen approach, which generates a degenerate two‑sheeted metric from a known vacuum solution and naturally leads to the regularized field equations. In Sec. 3 we define non‑degenerate and degenerate wormholes as two invariant classes under diffeomorphisms. Sec. 4 uses the thin‑shell and Klinkhamer metrics as concrete examples to show how both arise as solutions of a single \(g^{2}\)-modified Einstein–matter system and how the Einstein–Rosen bridge appears as a common limiting configuration. Sec. 5 summarizes the implications for NEC‑violation theorems and for the possible existence of stationary degenerate traversable wormholes without exotic matter.

\section{Two approaches to wormholes}

The division of wormholes into two types becomes natural if we consider the two main approaches used to describe wormholes. These are the Thorne approach \citep{MT}, which uses a \textit{non-degenerate} metric, and the Einstein-Rosen approach \citep{einstein2} , which uses a \textit{degenerate} metric. We describe these two approaches in more detail. 

\subsection{Thorne's approach}
 The starting point of Thorne's approach \cite{MT} is the choice of an arbitrary ansatz of a \textit{non-degenerate} two-sheeted metric describing the wormhole topology. Integral to this choice is the connection between global (two-sheeted) and local (one-sheeted) coordinates, providing a geometric interpretation of the metric in terms of a local observer. The metric determines the geometry (left-hand side of Einstein's equations), which simultaneously determines the stress-energy tensor that supports this geometry\footnote{The standard notations are adopted: \(g_{\mu\nu}\) is the metric tensor, \(R_{\mu\nu}\) and \(R\) are the Ricci tensor and its trace, \(T_{\mu\nu}\) is the stress-energy tensor; the relativistic system of units is used, in which the speed of light \(c\) and the gravitational constant \(G\) are equal to unity, \(c=G=1\).}
\begin{equation}
    R_{\mu\nu} - \frac{1}{2} g_{\mu\nu} R = 8\pi T_{\mu\nu}
\end{equation}
Thus, the problem boils down to choosing a suitable matter possessing precisely this stress-energy tensor. 

In this approach, the spherically symmetric Morris-Thorne metric, written in two-sheeted spherical coordinates \((t,l,\theta,\varphi)\), is typically used as a generalized wormhole ansatz:
\begin{equation}
    ds^{2}\big|^{\text{MT}} = -e^{2\Phi(l)}\, dt^{2} + dl^{2} + r^{2}(l)\, d\Omega^{2}.
\end{equation}
Here \(d\Omega^{2} = d\theta^{2} + \sin^{2}\theta\, d\varphi^{2}\) is the metric element of the solid angle. The two-sheeted coordinate \(l\) takes all values \(-\infty<l<+\infty\).  It is positive (\(l>0\)) on the first sheet and negative (\(l<0\)) on the second; at the boundary, i.e., at the throat, it vanishes \(l=0\). The two-sheet function \(r(l)\)  expresses the ordinary radial coordinate \(r\) in terms of the two-sheeted coordinate \(l\). Both functions, \(\Phi(l) \) and \(r(l)\), satisfy the natural conditions corresponding to flat space at infinity and are otherwise arbitrary.

Let us consider two specific examples of the application of Thorne's approach, in which the wormhole ansatz is analogous to the Schwarzschild metric.

\subsubsection{Thick shell ansatz (“smooth” wormhole)}
In this model, the relationship between the one-sheeted coordinate \(r\) and the two-sheeted coordinate \(l\) is given by
\begin{equation}
    r(l) = \sqrt{l^2 + b^2}
\end{equation}and the thick shell ansatz is\footnote{This metric is given in \cite{MT} for the case of zero mass parameter, \(M=0\). }
\begin{equation}
    ds^{2}\big|^{\text{thick-shell}} = -\left[1 - \frac{2M}{r(l)}\right] dt^{2} + dl^{2} + r^{2}(l)\,d\Omega^{2},
\end{equation}Here (as elsewhere below), the parameter \(b\), satisfying the condition \(b>2M\), has the meaning of the radius of the wormhole. The corresponding mixed \(tt\) component of the stress-energy tensor, which has the meaning of energy density, is expressed as
\begin{equation}
    T_t^t = -\frac{b^2}{8 \pi r^4} < 0.
\end{equation}is negative, i.e., it requires exotic matter.

The presence of matter can be seen by rewriting metric (4) in ordinary spherical coordinates:
\begin{equation}
   \begin{split}
 ds^2\bigg|^{\text{thick-shell}} = -\left(1 - \frac{2M}{r}\right) dt^2 + \left(1 - \frac{b^2}{r^2}\right)^{-1} dr^2 + r^2 d\Omega^2,\\ \quad (r > b).
   \end{split}
\end{equation}
This metric differs from the pure Schwarzschild metric in the radial component, indicating the presence of matter.

\subsubsection{Thin shell ansatz (“sharp” wormhole)}

In this model, two copies of the one-sheeted Schwarzschild space are connected through a sphere \(r=b\) \citep{visser2}. Then the relation between the one-sheeted coordinate \(r\) and the two-sheeted coordinate \(l\) has the form of a differential relation
\begin{equation}
    \frac{dr}{dl} = \pm \left[1 - \frac{2M}{r(l)}\right]^{1/2}
\end{equation}
with the initial condition \(r(0)=b\). This leads to the following Morris-Thorne metric
\begin{equation}
    ds^{2}\big|^{\text{thin-shell}} = -\left[1 - \frac{2M}{r(l)}\right] dt^{2} + dl^{2} + r^{2}(l)\,d\Omega^{2},
\end{equation}which, after substituting (7) into it, as it should be, turns into the Schwarzschild metric at all points in space outside the throat:
\begin{equation}
   \begin{split}
 ds^2\bigg|^{\text{thin-shell}} = -\left(1 - \frac{2M}{r}\right) dt^2 + \left(1 - \frac{2M}{r}\right)^{-1} dr^2 + r^2 d\Omega^2,\\ \quad (r > b).
   \end{split}
\end{equation} This metric corresponds to \(\delta\)-shaped energy density distribution,
\begin{equation}
    T_{t}^{t} = -\frac{\delta(l)}{2\pi b} \left(1 - \frac{2M}{b}\right)^{1/2} < 0,
\end{equation}which is concentrated at the throat \(r=b\), i.e., at \(l=0\). This energy density is negative, which again indicating the necessity of exotic matter in the case of a \textit{non-degenerate} metric.

Flamm paraboloids \citep{flamm} corresponding to thick- and thin-shell wormholes with metrics (4, 8) are shown in Fig. 1(a, b), see also Appendix A. The mass distribution of exotic matter in the thick-shell model, described by the spatial density (5), is shown in the first Flamm paraboloid, Fig. 1(a), in red of varying intensity. Exotic matter is smoothly distributed in space near the throat, depicted as a wide red "belt," whose color intensity decreases with distance from the throat. The mass distribution of exotic matter in the thin-shell model, described by the \(\delta\)-shaped distribution (10), is shown in the second Flamm paraboloid, Fig. 1(b), by a red circle coinciding with the equatorial section \( \theta = \pi /2\)  of the throat \(r=b\). This is an infinitely thin analog of the wide "belt" on the first paraboloid, showing the thick-shell case.

\begin{figure} [h]
 \centering
 \includegraphics[scale=0.425]{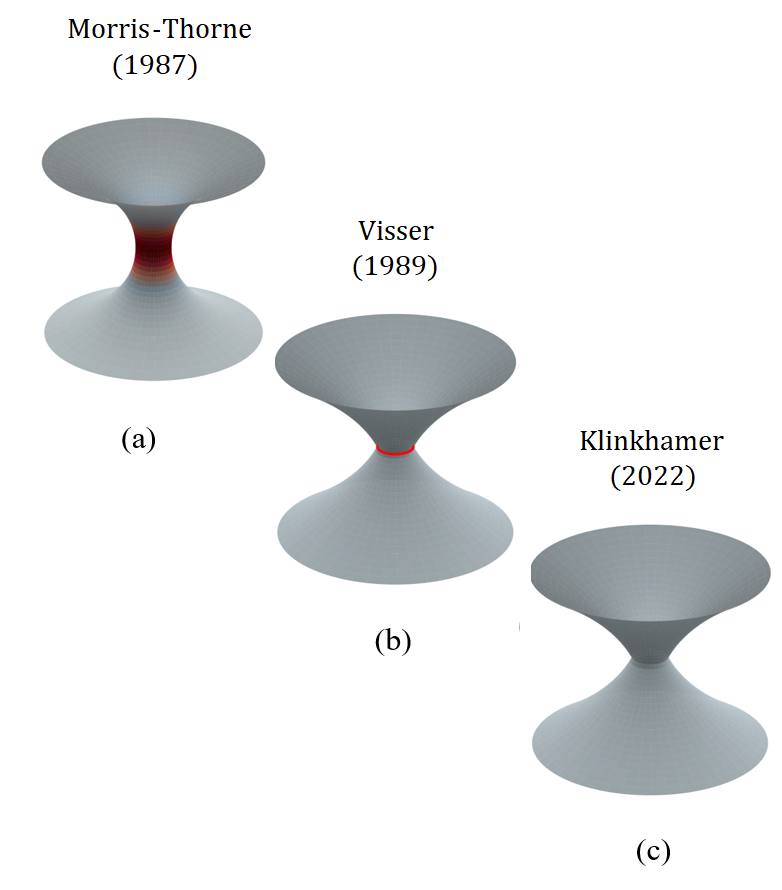}
  \caption{Historical development of traversable wormhole models: (a) Morris-Thorne thick-shell wormhole with distributed exotic matter (red region), (b) Visser thin-shell wormhole with exotic matter concentrated at the throat (red belt), (c) Klinkhamer degenerate wormhole in vacuum (no matter). Models (a) and (b) illustrate stationary \textit{non‑degenerate} wormholes with exotic matter; model (c) illustrates a \textit{degenerate} defect wormhole in vacuum described by the \textit{regularized} Einstein equations.}
\end{figure}

The two examples provided clearly demonstrate how Thorne's approach works (see Fig. 2). The approach is based directly on Einstein's equations with matter (1). By arbitrarily choosing a \textit{non-degenerate} metric \(g_{\mu\nu}\), describing a wormhole state, and substituting it into Einstein's equations, we are guaranteed to obtain the matter stress-energy tensor \(T_{\mu\nu}\) - which is the goal of this approach.
\begin{figure} [h]
 \centering
 \includegraphics[scale=0.44]{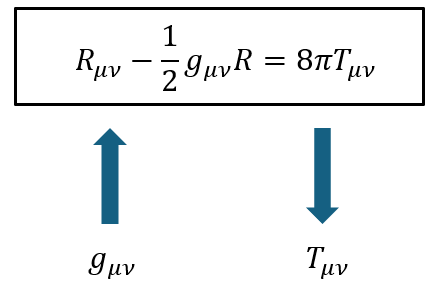}
  \caption{Flowchart of Thorne's approach.}
\end{figure}
Unfortunately, after this we are faced with the need for exotic matter, which does not yet allow us to move forward.

\subsection{Einstein-Rosen approach}

The Einstein-Rosen approach \citep{einstein2} differs fundamentally in its construction from Thorne's approach. The starting point of the Einstein-Rosen approach is a non-degenerate one-sheeted metric  \(g_{\mu\nu}\),. Unlike Thorne's approach, here the wormhole's two-sheeted metric is not chosen arbitrarily but is constructed based on some known solution \(g_{\mu\nu}\) of Einstein's equations in vacuum:
\begin{equation}
    R_{\mu \nu} = 0.
\end{equation}
The goal of the Einstein-Rosen approach is not to obtain a stress-energy tensor that satisfies Einstein's equations with matter and ensures the existence of a wormhole with a given metric.

The goal of the Einstein-Rosen approach is to obtain a new, two-sheeted metric \(\overline{g}_{\mu\nu}\), defined on a two-sheeted space with wormhole topology and satisfying Einstein's vacuum equations.

For this purpose, Einstein and Rosen proposed a procedure that implements a two-sheeted transformation of the radial coordinate \(r\). This transforms the one-sheeted radial coordinate \(r\) into a two-sheeted coordinate \(l\). As a result of this procedure, the one-sheeted manifold on which the original metric \(g_{\mu\nu}\) - the solution to Einstein's vacuum equations (11) - was defined - is transformed into a two-sheeted wormhole manifold.

The transition from the conventional radial coordinate \(r\) to the new, two-sheeted coordinate \(l\) is realized in the Einstein-Rosen approach by a certain two-sheeted differentiable function
\begin{equation}
    r = r(l).
\end{equation} The two-sheetedness of the transformation \(r(l)\) means that there exists a value of the radial coordinate \(r=B\) that corresponds to two different values of the two-sheeted coordinate \(l\); in ascending order, we denote them \(l_1\) and \(l_2\) (see Fig.3):
\begin{equation}
    r(l_1) = r(l_2) = B.
\end{equation}
\begin{figure} [h]
 \centering
 \includegraphics[scale=0.4]{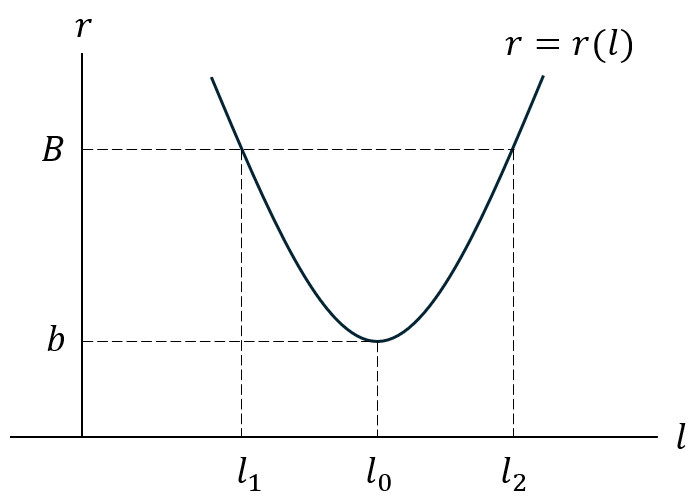}
  \caption{Graph of the function \(r(l)\).}
\end{figure}According to Rolle's theorem for differentiable functions, it follows that in the interval \((l_1, l_2)\), there exists a coordinate \(l=l_0\) for which the derivative \(r'(l)\) vanishes:
\begin{equation}
    r'(l_0) = 0
\end{equation}and the function \(r(l)\) itself reaches either a maximum or a minimum. In the case of a minimum (this case is shown in Fig. 3), we obtain a two-sheeted space with the wormhole topology; then, the quantity
\begin{equation}
    b = r(l_0)
\end{equation}
represents the radius of its throat. The Einstein-Rosen transformation thus transforms the original metric \(g_{\mu \nu}\) of a one-sheeted space, topologically equivalent to Minkowski space, into a new metric \(\overline{g}_{\mu\nu}\) of a two-sheeted space with the wormhole topology.

Due to the general covariance of Einstein's equations (11), the new metric \(\overline{g}_{\mu\nu}\), obtained from the original solution \(g_{\mu \nu}\) by coordinate transformation (12), must also be a solution of equations (11). However, it is precisely when attempting to substitute the new metric \(\overline{g}_{\mu\nu}\) into Einstein's equations (11) that we encounter a problem: this metric turns out to be \textit{degenerate}.

Indeed, in the radial part of the original metric \(g_{\mu \nu}\), the two-sheeted transformation (12) leads to the replacement
\begin{equation}
    dr^2 \rightarrow {r'}^2(l)\,dl^2.
\end{equation}
Since the derivative \(r'(l)\)  can vanish, this necessarily produces the degeneracy of the new (two-sheeted) metric \(\overline{g}_{\mu\nu}\). More specifically, according to (14) and (16), its determinant  \(\overline{g}(x) \equiv \det\|\overline{g}_{\mu\nu}(x)\|\) vanishes at the throat. In other words, after the Einstein-Rosen transformation, the original \textit{non-degenerate }one-sheeted metric  \(g_{\mu \nu}\) automatically turns into a \textit{degenerate} two-sheeted metric \(\overline{g}_{\mu\nu}\).

Since the expression for the curvature tensor contains the metric determinant in the denominator, this prevents us from calculating the curvature tensor at the throat and verifying that equations (11) hold for \(\overline{g}_{\mu\nu}\).

To address this issue, Einstein and Rosen proposed the following simple modification method \citep{einstein1} . It consists of multiplying Einstein's equations (11) by the factor \(g^2\), i.e., replacing equations (11) with the \(g^2\)-modified vacuum equations
\begin{equation}
    g^2  R_{\mu\nu} = 0.
\end{equation}
This \(g^2\)-modification eliminates the denominators in the Ricci tensor and renders it well-defined even when \(g=0\). We note that the full contraction \(g^{2}R=g^{2}g^{\mu \nu}R_{\mu \nu}\) would still contain one power of the inverse metric; it is therefore the component form  \(g^2  R_{\mu\nu} = 0\) — as used by Einstein and Rosen (1935a) — that provides the correct extended vacuum equations. More precisely, it eliminates the inverse metric  \(g_{\mu \nu}\) appearing in the denominators of the Christoffel symbols entering \(R_{\mu\nu}\), making each component \(g^2  R_{\mu\nu}\) a well-defined expression at \(g=0\). . Due to this, equations (17) are equally suitable for describing both the original \textit{non-degenerate} (one-sheeted) metric  \(g_{\mu \nu}\) and the new \textit{degenerate} (two-sheeted) metric \(\overline{g}_{\mu\nu}\). The modification expands the scope of application of the Einstein equations: the \(g^{2}\)-modified equations (17) not only have the same \textit{non-degenerate} solutions as the standard Einstein equations (11), but also admit other solutions expressed by \textit{degenerate} metrics. Moreover, due to the invariance of the factor \(g^2\) under coordinate transformations, the \(g^{2}\)-modified Einstein equations (17) remain covariant under coordinate transformations as the original Einstein equations (11).

Therefore, since the original one-sheeted metric \(g_{\mu \nu}\), being a solution to the standard Einstein equations (11), is guaranteed to also satisfy the \(g^{2}\)-modified Einstein equations (17), the new two-sheeted metric \(\overline{g}_{\mu\nu}\), obtained from \(g_{\mu \nu}\) by a coordinate transformation, is also guaranteed to satisfy the \(g^{2}\)-modified Einstein equations (17). In each specific case, this fact allows for a rather laborious verification; however, due to the covariance of equations (17), such a verification is redundant and not necessary.

Thus, the Einstein-Rosen approach achieves its goal of constructing a wormhole metric \(\overline{g}_{\mu\nu}\) satisfying the \(g^{2}\)-modified Einstein equations (17) in a vacuum\footnote{A generalization of the Einstein-Rosen approach to the Einstein-Maxwell equations and to a wormhole of arbitrary shape is presented in \citep{dimaschko}.}.  

As we noted above, see Eq. (16), the wormhole metric constructed in this way is automatically \textit{degenerate}. The flowchart of the Einstein-Rosen approach is shown in Fig.4
\begin{figure} [h]
 \includegraphics[scale=0.44]{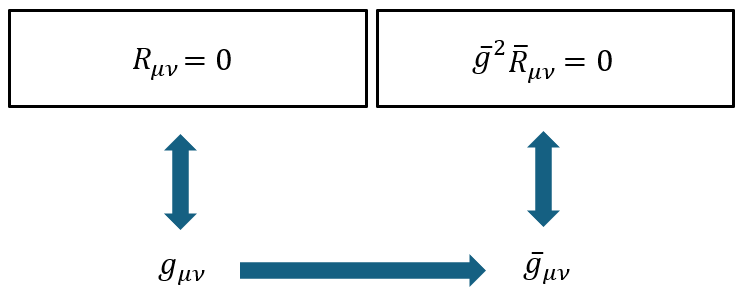}
 \centering
  \caption{Flowchart of the Einstein-Rosen approach. Here, the non-degenerate metric \(g_{\mu \nu}\) is the original solution of the standard Einstein equations in vacuum; using a two-sheeted coordinate transformation \(x^{\mu} \mapsto x^{\mu \prime}\), the non-degenerate metric \(g_{\mu \nu}\) generates a degenerate metric \(\overline{g}_{\mu\nu}\). The degenerate metric \(\overline{g}_{\mu\nu}\) automatically satisfies the \(g^{2}\)-modified Einstein equations in vacuum (18) and describes a two-sheeted state of space - a wormhole.}
\end{figure}

In their work, Einstein and Rosen took as the starting metric \(g_{\mu \nu}\) the Schwarzschild metric 
\begin{equation}
    ds^{2}\bigg|^{S} = -\left(1 - \frac{2M}{r}\right) dt^{2}
    + \left(1 - \frac{2M}{r}\right)^{-1} dr^{2}
    + r^{2} d\Omega^{2},
\end{equation}from which, by means of a two-sheeted transformation \(r=l^2+2M\), the first wormhole metric was obtained - the Einstein-Rosen bridge \citep{einstein2}.

Here, we consider as a concrete example another two-sheeted transformation of the Schwarzschild metric,
\begin{equation}
    r(l) = \sqrt{l^2 + b^2},
\end{equation}where the parameter \(b\), as in Thorne's approach, satisfies the condition \(b>2M\) and has the meaning of the wormhole radius. Transformation (19) transforms the Schwarzschild metric (one-sheeted and \textit{non-degenerate}) into the Klinkhamer metric \citep{klinkhamer1}.

\begin{subequations}
\begin{align}
ds^{2}\Big|^{K}
&= -\left(1-\frac{2M}{r}\right) dt^{2}
+ \left(1-\frac{2M}{r}\right)^{-1}
\frac{l^{2}}{r^{2}}\, dl^{2}
+ r^{2} d\Omega^{2}, \label{eq:20a} \\
r &= \sqrt{l^{2}+b^{2}}, \label{eq:20b}
\end{align}
\end{subequations}
which is two-sheeted and \textit{degenerate}. This metric satisfies the \textit{regularized} Einstein equations in vacuum, equation (17). The Klinkhamer metric (20) is a solution of the \(g^{2}\)-modified equations (17) in the entire two-sheeted space without exception, including the throat.\footnote{Specific arguments justifying the applicability of the degenerate metric on the wormhole throat within the framework of \(g^{2}\)-modified equations (17) are given in Appendix B.3 of the paper \citep{klinkhamer3}.} The \(g^{2}\)-modified equations (17) describe a nontrivial two-sheeted configuration. By general covariance of the Einstein equations, a two-sheeted coordinate transformation applied to a vacuum solution of the standard equations (11) yields a topologically nontrivial solution of (17) on the entire two-sheeted manifold, including at \(g = 0\). This is not a tautology but a consequence of covariance: the resulting metric is a specific degenerate solution of (17), rather than an arbitrary degenerate tensor field.
 
  \vspace{12pt}

 \fbox{\parbox{0.44\textwidth}{
 This is the first key point: critics claim that the Klinkhamer metric provides no information at the throat. However, this metric is a solution of the \(g^{2}\)-modified equations (17) everywhere, including at the throat, where \(g = 0\). Unlike the standard form (11), which becomes singular at \(g = 0\), the \(g^{2}\)-modified equations (17) remain well-defined and finite at the throat. Therefore, the Klinkhamer metric describes a regular state of the gravitational field throughout the entire two-sheeted spacetime.
}}

 \vspace{12pt}

In ordinary radial \(r\)-coordinates, the Klinkhamer metric has the form of the Schwarzschild metric with the constraint \(r \ge b\):
\begin{equation}
    ds^{2}\big|^{K} = -\left(1 - \frac{2M}{r}\right) dt^{2} + \left(1 - \frac{2M}{r}\right)^{-1} dr^{2} + r^{2} d\Omega^{2}, \quad r \geq b
\end{equation}
 Compared to the thin-shell metric (10), defined only for \(r>b\), the Klinkhamer metric also includes the throat, i.e., the surface \(r=b\). Outside the throat, the Klinkhamer metric coincides with the thin-shell metric (9); here, both metrics reproduce the Schwarzschild metric. The Klinkhamer metric and the thin-shell metric coincide outside the throat but differ at the throat.
 
 \vspace{12pt}
 
  \fbox{\parbox{0.44\textwidth}{
 This is the second key point. Outside the throat, the Klinkhamer metric coincides with the thin-shell metric (9); here, both metrics reproduce the Schwarzschild metric. At the throat itself, however, they are completely different: the thin-shell model requires a delta-function source with exotic matter (\(T_t^t<0\)), while the Klinkhamer metric satisfies the vacuum Einstein equations (\(T_{\mu}^{\nu}=0\)) in \(g^{2}\)-modified form. This discrepancy is not a manifestation of a coordinate defect but reflects a fundamental difference between \textit{non-degenerate} and \textit{degenerate} metrics. They describe two distinct classes of states within different forms of the Einstein equations .
}}
 
 \vspace{12pt}

Being a solution of the \(g^{2}\)-modified Einstein equations in a vacuum, the Klinkhamer metric describes the state of space without matter. This distinguishes it from the thin-shell metric (9), which is described by the standard (non-regularized) Einstein equations (1) and requires the presence of exotic matter.

The two-sheeted \textit{degenerate} Klinkhamer metric (20) is a self-consistent solution of the \textit{regularized} Einstein equations in a vacuum; in other words, here the only source of the gravitational field is the gravitational field itself. Einstein and Rosen \citep{einstein1} described their solution of the \(g^{2}\)-modified equations (17) as follows: “… which is the field-producing mass without requiring for this the introduction of any new field quantities.”

The Flamm paraboloid for the Klinkhamer metric is shown in Fig. 1(c). Since the Klinkhamer metric (21) outside the throat coincides with the thin-shell metric (9), their Flamm paraboloids outside the throat are also identical. The difference between them appears only at the throat: here, the thin-shell metric, described by standard (non-regularized) Einstein equations (1), requires exotic matter (the narrow red belt in Fig. 1(b)), while the Klinkhamer metric, described by \textit{regularized} Einstein equations (17), corresponds to a state without matter - a vacuum (no belt).

\section{Non-degenerate and degenerate wormholes}

The above examples give us grounds for dividing all wormholes into two alternative types:

1) \textit{Non-degenerate}, for which the determinant g(x) vanishes nowhere - for example, the Morris-Thorne wormholes (2), and

2) \textit{Degenerate}, for which the determinant \(g(x)\) vanishes at the throat - for example, the Einstein-Rosen bridge \citep{einstein2} or the Klinkhamer wormhole (20).

This division is invariant under transformations of reference frames. The reason for this invariance is that the transition between two physically realizable reference frames is always described by a diffeomorphism. This means that the Jacobian of the transformation \(J = \det \left\| \partial x^{\mu \prime}/ \partial x^{\nu} \right\|\) is non-zero and finite, i.e., \(J \ne 0\) and \(1/J \ne 0\). Since under the coordinate transformation \(x'=x'(x)\), the determinant g of the metric is transformed according to the law \(g'=gJ^{-2}\), a diffeomorphic transformation cannot change the very fact of equality/inequality to zero of the determinant \(g\). Therefore, under any transformation of the reference frame, a \textit{non-degenerate} wormhole remains \textit{non-degenerate}, just as a \textit{degenerate} wormhole remains \textit{degenerate}.

As follows from the comparative analysis in Section 2, the standard Einstein equations (1) are sufficient to correctly describe a \textit{non-degenerate} wormhole, while modified equations (17) must be used for a \textit{degenerate} wormhole. Therefore, appropriate methods for constructing \textit{non-degenerate} and \textit{degenerate} wormholes are, respectively, the Thorne approach (see Fig. 2) and the Einstein-Rosen approach (see Fig. 4).

\section{Heuristic analysis and a unified approach  }

A comparison of the Klinkhamer metric (21) with the metric of the thin-shell model (9), written in conventional spherical coordinates, reveals their complete similarity (this is clearly visible in their Flamm paraboloids, Fig. 1(b, c)). The difference appears only at the throat, where matter is present in the thin-shell model, while it is absent in the Klinkhamer wormhole. We will perceive this fact not as a contradiction. In fact, no contradiction exists, since these two metrics are solutions of two  different forms of the Einstein field equations, the standard form (1) and the \(g^{2}\)-modified form (17). Instead, we view this as an opportunity to trace, using this example, the similarity and difference between \textit{non-degenerate} and \textit{degenerate} metrics.  

To explore the analogy between these two metrics, we first consider the evolution of the thin-shell metric as the wormhole radius \(b\) decreases. Fig. 5(a) shows the evolution of the Flamm paraboloid of this metric for three successive values of \(b=6M\), \(b=4M\), and  \(b=2M\). As the throat radius \(b\) decreases, the energy density at it also decreases, as represented by the fading "red belt" around the throat. According to (11), this density is  \(\propto (1-2M/b)^{1/2} \) and vanishes at \(b=2M\). At this same point, the kink in the Flamm paraboloid at \(r=b\) disappears, and the paraboloid itself transforms into the classical "smooth" Flamm paraboloid of the Einstein-Rosen wormhole.

Next, we construct the same sequence of Flamm paraboloids for the Klinkhamer metric. Since the thin-shell metric (9) and the Klinkhamer metric (21) coincide outside the throat, their Flamm paraboloids will also be identical for the same values of \(b\). The difference from the thin-shell metric is that there is no matter at the throat of the Klinkhamer wormhole (there is no "red belt" for any value of  \(b\); see Fig. 5(b)).

\begin{figure} [h]
 \includegraphics[scale=0.75]{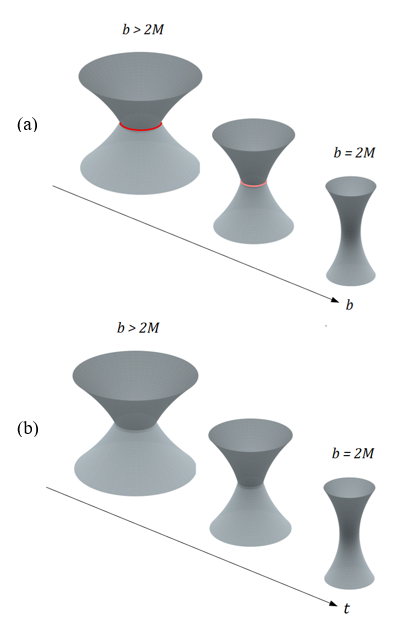}
 \centering
  \caption{Evolution of the Flamm paraboloid a) the thin-shell metric with decreasing wormhole radius \(b\); b) the Klinkhamer metric over time \(t\) during the gravitational collapse of the wormhole. In both panels the configuration with \(b=2M\) coincides with the Einstein–Rosen bridge, which acts as the common limiting state in the unified regularized description.}
\end{figure}

Is this the only difference?

No, it is not the only one. As shown in \citep{dimaschko1}, the Klinkhamer metric (20) with a constant \(b\) value, although a solution to the \(g^{2}\)-modified Einstein equations (17), does not lead to a stationary state of a degenerate wormhole. This is due to an additional topological degree of freedom - the throat position - which is absent in the one-sheeted space and determines the internal dynamics of the wormhole\footnote{The recognition that the wormhole throat constitutes an independent dynamical variable has precedent in the thin-shell wormhole literature \citep{visser2,visser}, where the throat radius enters as the configuration variable in a Wheeler-DeWitt quantization. The dynamical or statistical treatment of the topological variables has long been considered in discussions of topology change \citep{geroh,sorkin,horowitz,deborde}, wormhole formation \citep{hawking,visser,koga}, and topological quantum field theory \citep{atiyah,witten}. More recently, \citep{borissova} derived an effective action for throat dynamics from first principles, showing that the throat radius is 'a single degree of freedom' governed by its own equation of motion derived from the Israel junction formalism. \citep{dimaschko1} extended this concept to degenerate wormholes, where the standard Einstein equations are insufficient and the extended equivalence principle determines the throat dynamics. }.  As a result, the Klinkhamer wormhole acquires the meaning of a dynamical wormhole with a uniquely determined dependence of the wormhole radius on time, \(b(t)\). This dependence represents the radial dynamics of the Klinkhamer wormhole, in particular, its collapse under the influence of its own gravitational field. During collapse, the Klinkhamer wormhole passes through states described by metric (20) with successively decreasing values of \(b\).

In Fig. 5(a, b), this difference is indicated by the fact that in case (a) the evolution of the wormhole is shown along the axis of the parameter \(b\), and in case (b) – along the time axis \(t\). Since in both cases the final state of the evolution is the Einstein-Rosen wormhole, this state serves as a bridge connecting \textit{non-degenerate} (in this case, thin-shell) and \textit{degenerate} (in this case, Klinkhamer’s) wormholes. In the first case, it is one of many stationary states of a \textit{non-degenerate} wormhole, having a minimum radius \(b=2M\) and free of matter\footnote{In the case of a thin-shell wormhole, the limit \(b \rightarrow 2M\)  leads to only the energy density vanishing, while the pressure diverges, see  (\citep{visser}, Eqs. (15.45–46)). Therefore, the stress-energy tensor as a whole does not vanish. This is an artifact of the model and cannot change the fact that at \(b=2M\) we arrive at the Einstein-Rosen wormhole, where matter is absent.}; in the second case, it is the only possible stationary state of a \textit{degenerate} wormhole, which is the limiting point of its evolution.

A comparison of the parametric evolution of a \textit{non-degenerate} wormhole and the time evolution of a \textit{degenerate} wormhole yields a natural picture of radial dynamics with a single stable equilibrium state—the Einstein-Rosen wormhole. In the absence of a matter factor (stress-energy tensor), the \textit{degenerate} wormhole collapses into this equilibrium state (\(b=2M\)). In the presence of a matter factor, the wormhole can be stabilized in another, \textit{non-degenerate }state (\(b>2M\)); typically, such a factor is exotic matter. Since these two regimes are currently described by different Einstein equations—standard (1) and \(g^{2}\)-modified (17)—this unified picture suggests the need for their joint description within a unified approach. This description should a) include the possibility of the presence of matter, and b) encompass the sector of \textit{degenerate} metrics, allowing for \(g=0\). Consequently, a natural form of such a unified approach is the \(g^2\)-modified Einstein equations with matter:
\begin{equation}
    g^2 \left( R_{\mu\nu} - 8\pi T_{\mu\nu} + 4\pi g_{\mu\nu} T \right) = 0
\end{equation}
where \(T = g_{\mu\nu} T^{\mu\nu}\) is the trace of the stress-energy tensor. This form follows directly from the trace-reversed standard Einstein equations  \(R_{\mu\nu} = 8\pi \left( T_{\mu\nu} - \frac{1}{2} g_{\mu\nu} T \right)\) multiplied by \(g^2\), and avoids the Ricci scalar \(R\) whose contraction would reintroduce a factor of the inverse metric. 

 Equations (22)  represent a unified \(g^{2}\)-modified Einstein–matter system in which non‑degenerate thin‑shell wormholes and degenerate defect wormholes appear as two qualitatively distinct classes of states within the same underlying theory. Any solution of the standard Einstein equations with matter (1), as well as the \(g^{2}\)-modified Einstein equations in vacuum (17), is a solution of the system (22).\footnote{Equation (22) admits a variational derivation from the polynomial action of \citep{peres} and \citep{katanaev}, which replaces the Einstein–Hilbert Lagrangian \(\sqrt{-g}R\) by a polynomial expression in the metric components and their derivatives; standard GR is recovered in the non-degenerate sector \(g \ne 0\) . Whether (22) should be regarded as a fundamental equation or as an effective description depends on one's perspective: within the extended configuration space that includes degenerate metrics it plays the role of the fundamental field equation, while standard GR arises as its restriction to \(g \ne 0\) } The converse, generally speaking, is not true.

In particular, both the \textit{non-degenerate} metric (8) of the thin-shell model and the \textit{degenerate} Klinkhamer metric (20) are solutions of the same regularized system of Einstein equations with matter (22). Therefore, these equations satisfy all the stated conditions. It is worth noting that these equations furthermore allow for \textit{degenerate states with matter} that go beyond the traditional Thorne and Einstein-Rosen approaches; this may provide a new direction of research.

We note that the present discussion uses the \(g^2\)-modification following \citep{einstein1} and \citep{katanaev}, which is sufficient to describe the vacuum and thin-shell configurations considered here. A more general \(g^k\)-modification, with \(k>2\), has recently been argued to be necessary for certain matter components \citep{klinkhamer4}; the implications of such a generalization for the unified framework presented here are left for future work. 

\section{Conclusions and outlook }

Thus, we have introduced the notion of a degenerate wormhole, characterized by a vanishing metric determinant g on the throat and described by the polynomial, \(g^{2}\)-modified form of Einstein’s equations. This extends the usual picture based on the standard Einstein equations, which by construction only access the non‑degenerate sector of wormhole geometries. Both sectors can be treated within a single unified framework by employing the \(g^{2}\)-modified Einstein equations with matter.

We have shown that the Thorne approach, built on a two‑sheeted Morris–Thorne ansatz and the standard Einstein equations with matter, always leads to non‑degenerate wormholes supported by exotic stress–energy. In contrast, the Einstein–Rosen approach, which generates a two‑sheeted geometry from a known vacuum solution and uses the \(g^{2}\)-modified equations, naturally produces degenerate wormholes in vacuum, with the Einstein–Rosen and Klinkhamer metrics as explicit examples.

A detailed comparison between the non‑degenerate thin‑shell wormhole and the degenerate Klinkhamer wormhole shows that they represent genuinely distinct classes of states rather than different coordinate descriptions of the same configuration. The thin‑shell wormhole is a stationary non‑degenerate solution in the presence of exotic matter, while the Klinkhamer configuration is a dynamical degenerate vacuum wormhole that collapses towards the Einstein–Rosen bridge as a limiting state. The Einstein–Rosen wormhole thus appears as a common endpoint: a purely mathematical limit \( b\to 2M\) for the thin‑shell family and the late‑time limit of the degenerate collapse.

This unified viewpoint resolves the apparent contradiction between thin‑shell and defect wormholes: they arise as two distinct classes of states—non‑degenerate with matter and degenerate vacuum—following from the same \(g^{2}\)-modified Einstein–matter equations. It also clarifies that classical no‑go theorems based on NEC violation are tied to the non‑degenerate sector and do not automatically extend to degenerate wormholes. In this light, the existence of stationary degenerate traversable wormholes that respect the NEC becomes a concrete question for further investigation within the \(g^{2}\)-modified framework.

\appendix

\section{Flamm's paraboloids}

Here we construct Flamm paraboloids for the wormhole metrics of interest to us. 

The general definition of Flamm paraboloid \citep{MTW} for an arbitrary spherically symmetric metric follows from the local Pythagorean relation, which expresses the square of the differential of the proper radial coordinate \(dl^2\) through the sum of the squares of the differentials of the ordinary \(r\)-coordinate \(dr^2\) and the auxiliary \(z\)-coordinate \(dz^2\):
\begin{equation}
    dl^2 = dr^2 + dz^2
\end{equation}For a stationary spherically symmetric metric, the square of the differential of the proper radial coordinate is
\begin{equation}
    dl^2 = g_{rr}(r)\,dr^2,
\end{equation}which yields the differential equation
\begin{equation}
    dz = \pm \left[ g_{rr}(r) - 1 \right]^{1/2} \, dr.
\end{equation}
Its solution, expressed by the function \(r(z)\), determines the shape of the Flamm paraboloid in cylindrical coordinates \((r,\varphi ,z)\). The metric induced in Euclidean space on this paraboloid is identical to the spatial metric of the equatorial section \(\theta = \pi /2 \) of the wormhole.

We are interested in two cases:

 1) the thick-shell model, when, according to (6),
\begin{equation}
    g_{rr}(r) = \left( 1 - \frac{b^2}{r^2} \right)^{-1},
\end{equation}

 2) the thin-shell model, when, according to (9),
\begin{equation}
    g_{rr}(r) = \left(1 - \frac{2M}{r}\right)^{-1}.
\end{equation}
For the unique determination of the solution \(r(z)\), differential equation (A.3) requires one boundary condition, which we choose as
\begin{equation}
    r(0) = b.
\end{equation} Its geometric meaning is that the throat, with radius b, corresponds to a section of the paraboloid by the plane \(z=0\). Given this condition, for the two thick-shell and thin-shell models of interest to us, the function r(z) has the form, respectively,
\begin{equation}
    r(z) = b \cosh\left(\frac{z}{b}\right) \qquad \text{(thick shell)}
\end{equation}

\begin{equation}
    r(z) = b + \frac{z^2}{8M} + \lvert z \rvert \sqrt{\frac{b}{2M} - 1} \quad \text{(thin shell)}
\end{equation}

Since the thin-shell model metric (9) is identical to the Klinkhamer metric (21) outside the throat, the latter expression also describes the shape of Flamm’s paraboloid for the Klinkhamer metric.

\section*{Declaration of Competing Interest}

The author declares that he has no known competing financial interests or personal relationships that could influence the content of this paper.

\section*{Data availability}

No data was used for the research described in the article.

\bibliographystyle{elsarticle-harv} 
\bibliography{main}



%% For example:

%% \bibitem[Aladro et al.(2015)]{Aladro15} Aladro, R., Martín, S., Riquelme, D., et al. 2015, \aas, 579, A101

%%\end{thebibliography}

\end{document}